# Impact of gate voltage on switching field of perpendicular magnetic tunnel junctions with a synthetic antiferromagnetic free layer

K. Fan*[1,2], S. V. Beek[1], G. Talmelli[1], V. Kateel[1], D. Giuliano[1,3], B. Vermeulen[1,3], K. Cai[1], B. Sorée[1,2,4], J. D. Boeck[1,2], R. Carpenter[1], S. Rao[1], S. Couet[1], V. D. Nguyen[1]* and G. S. Kar[1]

[1]*Interuniversity Microelectronics Center (IMEC), Kapeldreef 75, 3001 Leuven, Belgium*
[2]*Department of Electrical Engineering, ESAT-INSYS Division, Katholieke Universiteit Leuven, Kasteelpark Arenberg 10, 3001 Leuven, Belgium*
[3]*Department of Physics and Astronomy, QSP Division, Katholieke Universiteit Leuven, Celestijnenlaan 200D Box 2414, 3001 Leuven, Belgium*
[4]*Department of Physics, Universiteit Antwerpen, Groenenborgerlaan 171, 2020 Antwerp, Belgium*

*\* Corresponding authors: Kaiquan.Fan@imec.be and Van.Dai.Nguyen@imec.be*

**Abstract**

We present micromagnetic simulations and experiments on voltage-assisted field switching in perpendicular magnetic tunnel junctions (MTJs) with a synthetic antiferromagnetic (SAF) free layer, where the magnetic state of one sublayer is detected via tunneling magnetoresistance (TMR). Simulations reveal that local modulation of perpendicular magnetic anisotropy (PMA) in one SAF sublayer leads to distinct switching characteristics. The switching field varies linearly with the anisotropy field, indicating voltage-controlled magnetic anisotropy (VCMA)-dominated dynamics similar to single free-layer devices. We then experimentally study the magnetic switching field of MTJ devices with SAF free layers under applied gate voltage. By varying the MgO tunnel barrier thickness to systematically modulate the resistance-area (RA) product, we enable quantitative separation of spin-transfer torque (STT), VCMA, and Joule heating contributions. Our findings indicate that VCMA dominates in devices with a high RA product, while low-RA devices exhibit nonlinear switching behavior due to enhanced contributions from STT and Joule heating. Furthermore, the effective fields derived from STT, VCMA, and Joule heating contributions under various gate voltages show minimal dependence on device critical dimensions, indicating favorable scaling behavior. This work presents a unified framework analyzing the roles of STT, VCMA, and Joule heating in SAF-based voltage-gated spin-orbit torque (SOT) MRAM, offering key insights for the optimization of performance, energy efficiency, and scalability in SOT-MRAM technologies.

Magnetic random-access memory (MRAM) is a promising candidate for embedded memory due to its non-volatility, speed, low power, and CMOS compatibility.[1–3] Among MRAM technologies, spin-orbit torque (SOT)-MRAM, with separate read and write paths, addresses the speed-endurance trade-offs of spin-transfer torque (STT)-MRAM,[4–7] offering a fast, high-endurance, and reliable alternative to SRAM replacement.[8–10] However, the 3-terminal design of SOT-MRAM requires at least two transistors per memory cell, limiting density capability. Voltage-gated SOT-MRAM (VGSOT-MRAM) has been proposed to mitigate this limitation by utilizing the voltage-controlled magnetic anisotropy (VCMA) effect to reduce the SOT switching current, also enabling selective writing in high-density multi-pillar architectures.[11,12] This concept has been demonstrated using CoFeB-based free-layer magnetic tunnel junctions (MTJs) grown on β-W-based SOT tracks.[11,13] However, integration into back-end-of-line (BEOL) processes with annealing temperatures up to 400 °C remains challenging due to interfacial material intermixing and degradation of perpendicular magnetic anisotropy (PMA).[14] Synthetic antiferromagnetic hybrid free layer (SAF-HFL) has been proposed as a promising alternative to overcome these issues,[14–16] offering fast switching, enhanced retention, and tunable magnetic properties through modulation of interlayer exchange coupling and anisotropy in each sublayer.[17,18] Therefore, integrating VG-SOT with SAF-HFL is expected to reduce power consumption and enable higher integration density for next-generation memory technologies.

Previous studies reported that applying a gate voltage across the MTJ induces STT, VCMA, and Joule heating effects, which collectively influence the switching of the CoFeB free layer.[19,20] In particular, the VCMA effect directly modulates the magnetic anisotropy linearly, thereby altering the switching.[11,20,21] However, SAF structures, comprising two ferromagnetic layers antiferromagnetically coupled via a Ru spacer, exhibit more complex switching behavior, governed by the coupling strength and the relative magnetic anisotropy of the two layers.[17,22,23] Therefore, it remains unclear how voltage-induced local modulation of anisotropy in one sublayer impacts the overall switching behavior of the SAF-HFL system. Additionally, most of the previous research has focused on MTJs with a low resistance-area (RA) product, typically below 100 Ω·µm².[24,25] In such devices, thin MgO barriers permit considerable currents to flow through, which can degrade the barrier and impact device reliability.[26,27] This high tunnelling current also generates pronounced STT and Joule heating effects, often overshadowing the contribution of the VCMA effect.[24] Consequently, the complex interplay of STT, VCMA, and Joule heating in low-RA devices incorporating SAF-HFLs makes their individual contributions difficult to disentangle.

This work investigates the influence of gate voltage on the switching dynamics of SAF-HFL structures through combined micromagnetic simulations and experiments. The simulations reveal that local modulation of PMA within the SAF sublayers leads to distinct switching modes, while the switching field exhibits a linear dependence on the anisotropy field $B_k$, consistent with VCMA behavior in single free-layer systems. Experimentally, by varying MgO tunnel barrier thicknesses, we further model and quantify the competing contributions of STT, VCMA, and Joule heating by analyzing the evolution of the hysteresis loop under varying gate voltages. Additionally, we investigate the dependence of these effects on the RA products and MTJ dimensions to isolate their individual contributions. These results advance the understanding of voltage-controlled magnetization dynamics in coupled magnetic systems and support the development of SAF-HFL-based VGSOT-MRAM devices.

We investigate three-terminal perpendicular magnetized top-pinned MTJs with a Pt SOT track and a SAF-based free layer as illustrated in Fig. 1(a). The SAF-HFL comprises a multilayer stack of CoFeB (0.8nm)/Spacer 1 /Co 1 (0.8nm)/Spacer 2 /Co 2 (0.8nm) where Spacer 1 and Spacer 2 correspond to WCoFeB and Ru consistent with our previous study, respectively.[28,29] The top CoFeB layer is ferromagnetically coupled with the top Co1 layer, forming the composite layer referred to as M1. Meanwhile, this M1 layer couples antiferromagnetically with the bottom Co 2 layer via Ruderman-Kittel-Kasuya-Yosida (RKKY) interaction.[30–32] To systematically investigate the effects of gate voltage across the MgO tunnel barrier, devices with varying RA products were fabricated by tuning the MgO thickness to 1.3 nm, 1.6 nm, and 1.7 nm, corresponding to RA values of 110, 600, and 1500 Ω·μm², respectively. All samples were annealed at 400 °C for 30 minutes. These SOT-MRAM devices were fabricated and integrated by imec's 300 mm MRAM platform as reported in previous studies.[14,15,17]

In our experiments, a sensing voltage is applied through the MTJ's top electrode to measure resistance, therefore M1 layer switching is directly monitored via tunnel magnetoresistance (TMR).[33] This sensing voltage also serves as a gate voltage, dropping mainly across the MgO barrier to modulate the effective PMA of the M1 layer linearly via the VCMA effect.[11] Unlike a conventional single free layer, the SAF device introduces additional complexity due to the interplay of coupling and anisotropy among two magnetic layers. As previously reported, the switching characteristics of SAF can display a two-step or three-step process depending on the exchange coupling strength between the two magnetic layers and their relative magnetic properties.[34,35]

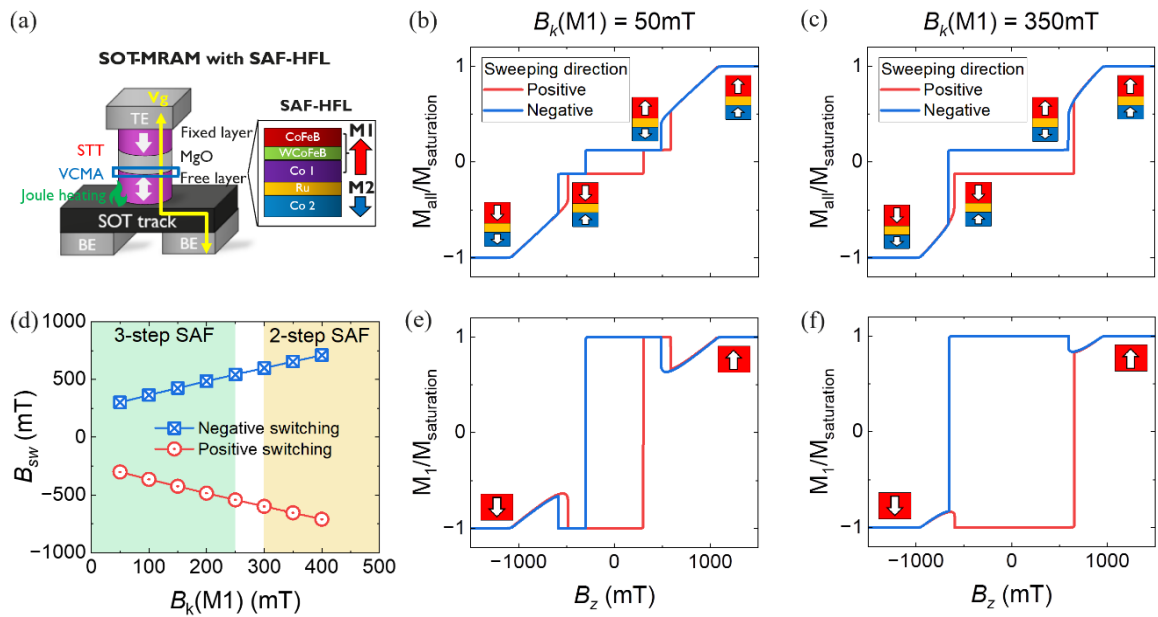

Figure 1(a) The schematic of STT, VCMA, and Joule heating effect in VGSOT-MRAM devices, including the schematic of SAF-HFL. The simulated hysteresis loop of all layers when (b) $B_k$ (M1) = 50mT (c) $B_k$ (M1) = 350mT. (d) The summarized $B_{sw}$ as a function of $B_k$ (M1). The simulated hysteresis loop of M1 layer when (e) $B_k$ (M1) = 50mT (f) $B_k$ (M1) = 350mT.

To understand how local modulation of anisotropy in the M1 layer affects the switching fields of the SAF-HFL system, we performed micromagnetic simulations using MuMax3.[36] A simplified two-layer SAF model was adopted to represent the SAF-HFL stack, comprising an M1 layer (composite of CoFeB/Co 1) and an M2 layer (representing Co 2). The detailed simulation parameters are provided in supplementary materials. We hypothesized that the gate voltage tuned the PMA of the M1 layer due to the voltage drop across the CoFeB/MgO interface. Accordingly, the effective anisotropy field of M1, $B_k$(M1), was varied from 50 mT to 400 mT, while $B_k$(M2) was fixed at 200 mT. Figures 1(b) and 1(c) show the simulated hysteresis loops of the SAF structure under out-of-plane field sweeps for two representative values of $B_k$(M1). When the PMA of M1 is weak, the SAF exhibits a three-step switching process, characterized by the concurrent switching of M1 and M2. During the positive field sweep (i.e., red curve of Fig.1(b)), an initial switching event is observed prior to crossing zero field, corresponding to the activation of the SAF coupling. With increasing field, a concurrent reversal of both magnetic layers occurs, followed by the SAF saturation regime at higher fields. In contrast, under a strong PMA of M1, a two-step reversal is observed, corresponding to activation and saturation of SAF configuration. In our device, the switching of the M1 layer is electrically detected via the TMR signal. Therefore, Fig. 1(d) summarizes the evolution of the M1 switching fields as a function of $B_k$(M1), while Figs. 1(e) and 1(f) show the corresponding hysteresis loops, enabling direct comparison with experimental results later. Compared to the hysteresis loop of the M1 layer (cf. Fig. 1(e) and 1(f)) and the full SAF structure $M_{all}$ (cf. Fig. 1(b) and 1(c)), the M1 switching occurs concurrently during the three-step SAF switching but during SAF saturation in the two-step SAF. Notably, regardless of the switching SAF characteristics, the switching field of M1 shows a linear dependence on $B_k$, consistent with VCMA behavior observed in single free-layer systems.

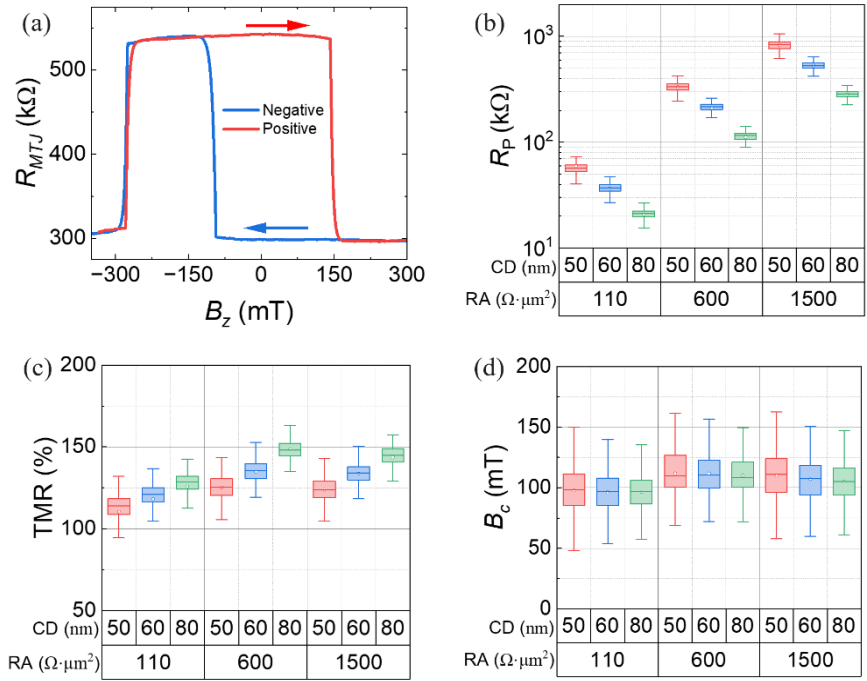

*Figure 2. (a) TMR-based hysteresis loop readout of SAF-HFL in SOT-MRAM devices. Box plots of (b) MTJ resistance in the parallel state. (c) TMR and (d) Coercivity as functions of the RA product (Ω·μm²) and CD (nm). The central line in each box represents the median, while box edges correspond to the first and third quartiles. Whiskers indicate the range within 1.5× the interquartile range. The filled circle denotes mean.*

To validate the simulation predictions, SAF-HFL-based MTJ devices with varying RA products and critical dimensions (CDs) were characterized, based on statistical data from 300 devices. Fig. 2(a) exhibits a typical TMR hysteresis loop of 80 nm MTJ device, capturing the magnetization reversal of the M1 layer. The parallel-state resistance ($R_p$) in Fig. 2(b) exhibits a decrease with larger CDs, which is consistent across all RA categories. An enhancement of $R_p$ for higher RA devices is attributed to increased MgO barrier thickness, which reduces tunneling current and raises $R_p$. The median TMR ratio in Fig. 2(c) is approximately 120% for devices with RA = 110 Ω·μm². The TMR ratio slightly increases with increasing RA, which can be attributed to an improved CoFeB/MgO interface quality associated with thicker MgO barriers.[15] The coercivity field ($B_c$) in Fig. 2(d) increases slightly from ~100 mT to ~110 mT with increasing RA. This trend reflects enhanced PMA in devices with thicker MgO, consistent with better CoFeB/MgO interface. Moreover, from the switching field distributions over 400 switching cycles, a thermal stability factor of $\Delta \approx 120$ is extracted in 80 nm CD devices, indicating adequate retention (see Supplementary Materials S2).

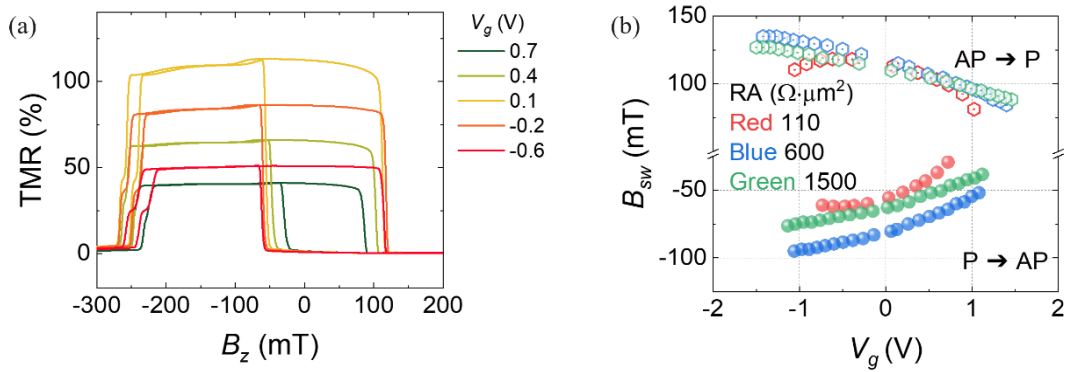

Figure 3. (a) TMR hysteresis loops under varying gate voltages in an RA 110 Ω·μm² device. (b) Summarized free layer switching fields as functions of gate voltages among devices with different RAs.

To examine the voltage-gate effects on SAF-HFL switching, we measured the TMR hysteresis loops of SAF-HFL devices under varying gate voltages and extracted switching fields. Fig. 3(a) shows representative TMR hysteresis loops under varying gate voltages for an RA = 110 Ω·μm² device with CD = 80 nm. Averaged over 50 cycles, the square loops and high TMR indicate robust PMA and consistent device quality across all voltages. A significant modulation of the switching field with gate voltage is observed, confirming tunable free layer magnetic properties. Fig. 3(b) summarizes the switching fields for both P→AP and AP→P transitions as functions of gate voltage for devices with different RA values. In RA = 600 and 1500 Ω·μm² devices, the switching fields exhibit a linear dependence on gate voltage[11,20], in agreement with the simulation results. In contrast, RA = 110 Ω·μm² devices show a parabolic modulation, indicating additional contributions from STT and Joule heating, which become significant in low-RA devices due to increased tunneling current.[19,24]

To quantify the individual contributions from each effect, we express the switching field as a function of gate voltage using the following model[19]

$$\begin{cases} B_{sw}^{P \to AP} = B_{c0} + B_{off} + \alpha \dfrac{V_g}{RA(V_g)} + \epsilon V_g - \zeta \dfrac{V_g^2}{RA(V_g)} \\ B_{sw}^{AP \to P} = -B_{c0} + B_{off} + \alpha \dfrac{V_g}{RA'(V_g)} - \epsilon V_g + \zeta \dfrac{V_g^2}{RA'(V_g)} \end{cases} \quad (1)$$

where the switching fields during both the P to AP and AP to P processes are considered simultaneously. Here, $B_{c0}$ represents the intrinsic coercivity of the free layer, and $B_{off}$ denotes the stray field from the reference and hard layers. The coefficients $\alpha, \epsilon$ and $\zeta$ describe the contributions from STT, VCMA, and Joule heating effects, respectively. RA and RA' are the resistance-area products under the P and AP states.

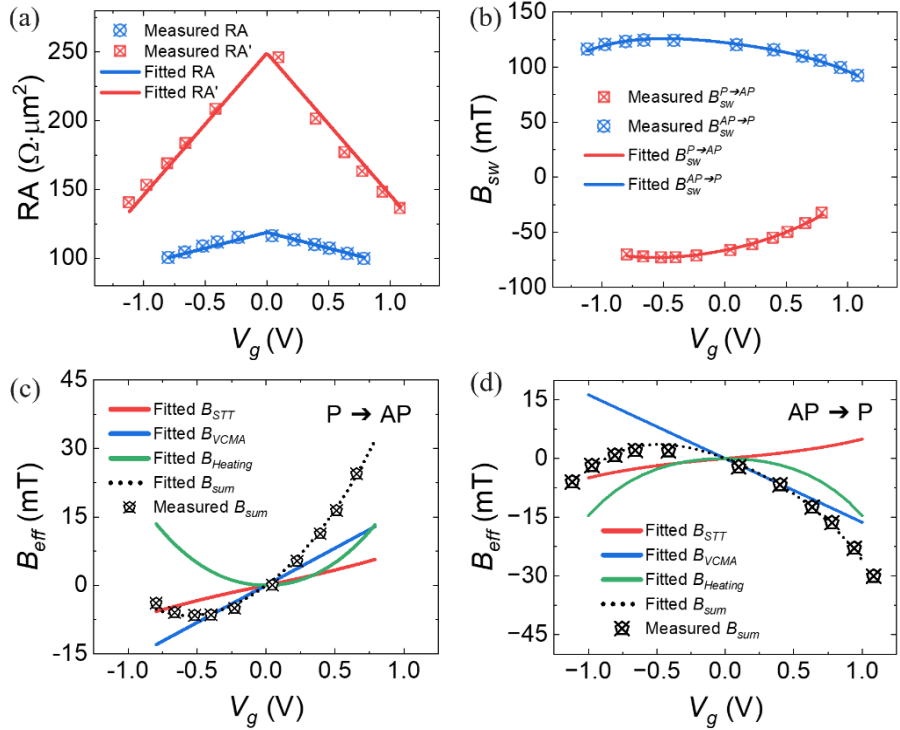

Figure 4. In an RA 110 Ω·μm² device, gate voltage dependence of measured and fitted (a) RA and RA'. (b) $B_{sw}$ during P to AP process and AP to P process. Fitted STT (fitted $B_{STT}$), VCMA (fitted $B_{VCMA}$), and Joule heating (fitted $B_{Heating}$) effective fields with the fitted sum (fitted $B_{sum}$) of all effective fields and the measured sum (measured $B_{sum}$) in (c) P to AP process (d) AP to P process.

In this analysis, we first examined the voltage dependence of MTJ resistance. As shown in Fig. 4(a), in an RA 110 Ω·μm² and CD 80 nm device, both the RA in the P state and RA' in the AP exhibit a clear dependence on gate voltage. This behavior is described by the following expression:

$$\begin{cases} RA(V_g) = RA_0 + a|V_g| \\ RA'(V_g) = RA'_0 + b|V_g| \end{cases} \quad (2)$$

where $RA_0$ and $RA_0'$ are fitted RA under zero gate voltage for the P and AP states, respectively. For each device, $RA(V_g)$ and $RA'(V_g)$ were globally fitted using Equation 2. These fits were then substituted into Equation 1 to perform a second global fit for the switching fields $B_{sw}^{P \rightarrow AP}$ and $B_{sw}^{AP \rightarrow P}$, capturing the combined influence of all three effects. Fig. 4(b) highlights the switching fields for transitions between P and AP states and vice versa. The solid fitted curves align well with the experimental data, validating the analytical framework. As shown by simulation above, different SAF configurations exhibit distinct switching pathways including two-step or three-step switching characteristics. However, the switching field of M1 displays a linear dependence on $B_k$ across all cases. This indicates that this unified model can be extended to different SAF designs, as variations in SAF do not affect the underlying physical mechanisms captured by the model.

To quantitatively compare STT, VCMA, and Joule heating effects under various gate voltages, each effect was converted into an effective field using Equation 1. Figs. 4(c) and (d) show these calculated effective fields as functions of gate voltage during field-driven switching, including the calculated sum of all effective fields (*Fitted $B_{sum}$*) and measured sum (*Measured $B_{sum}$*). The measured sum was determined as the measured switching field minus the constant intrinsic coercivity and offset field for each specific device. The STT effective field ($B_{STT}$) exhibits an approximately linear dependence on $V_g$, with an approximate slope of 7 mT/V. This near-linear trend originates from the voltage-induced modulation in MTJ resistance, which directly affects the current density. Despite its linear increase, the contribution of STT remains smaller than that of VCMA and Joule heating effects at higher voltages, indicating its secondary role in these conditions. The VCMA effective field ($B_{VCMA}$) demonstrates a linear dependence on $V_g$, with a higher slope of 16 mT/V. This effect is decided by the voltage polarity, aiding switching effectively when the gate voltage is positive, consistent with modulation of anisotropy at the CoFeB/MgO interface. This observed voltage-induced modulation corresponds to an effective SAF response, while the intrinsic VCMA contribution is expected to be comparable to that of conventional CoFeB/MgO-based devices, owing to the presence of a similar CoFeB/MgO interface. We compared gate voltage impact on effective magnetic anisotropy field between SAF-HFL and single free layer in supplementary material S4. The Joule heating effective field ($B_{Heating}$) shows a symmetric dependence on $V_g$, peaking at approximately 25 mT for $|V_g|$=1.0 V. This symmetry reflects the quadratic dependence of Joule heating on voltage, as the dissipated power scales with $V_g^2$ and is therefore independent of polarity. As a result, the Joule heating contribution increases rapidly at higher gate voltages, effectively reducing the switching energy barrier.

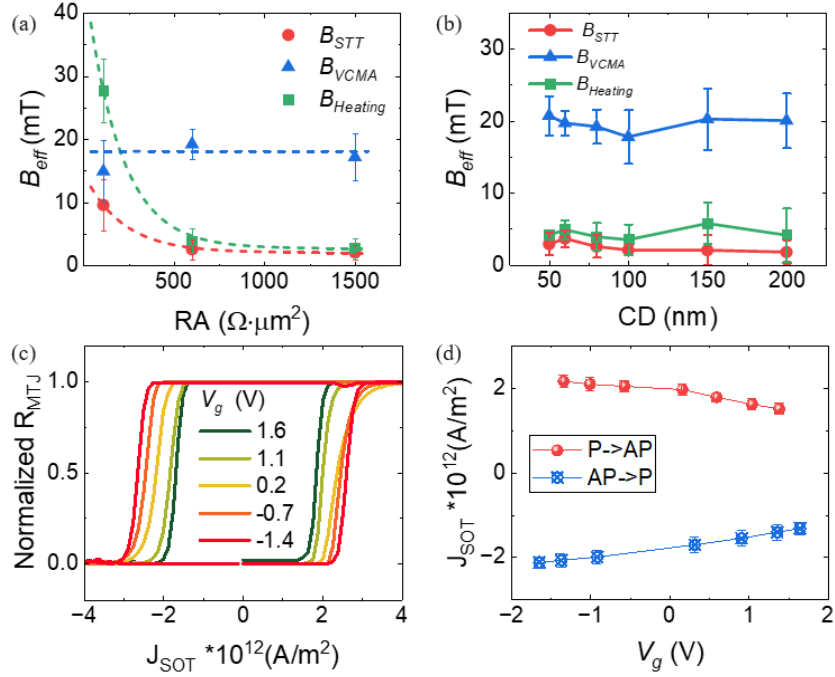

Figure 5 (a) RA dependence of effective fields at $V_g$ = 1 V for 15 CD = 80 nm devices, the dashed lines as guide for the eye. (b) CD dependence of effective fields at $V_g$ = 1 V for 15 RA = 600 $\Omega \cdot \mu m^2$ devices. (c) SOT switching loops under varying gate voltage with 1ns SOT pulse and $B_x$=40mT. (d) Summarized critical SOT switching current density as a function of gate voltage.

| RA ($\Omega \cdot \mu m^2$) | VCMA Voltage (V) | STT Current (μA) | Current Density (*$10^9 A/m^2$) | Joule heating (°C) |
|---|---|---|---|---|
| 110 | 1 | 54.0 | 1.8 | 33.8 |
| 600 | 1 | 9.5 | 1.9 | 4.8 |
| 1500 | 1 | 4.1 | 0.8 | 3.4 |

Table 1. Estimated STT current, current Density, and device temperature increase under 1 V gate voltage for different RA values.

To evaluate the scalability of VGSOT devices, we investigate the RA and CD dependence of all effective fields under 1V gate voltage from 15 devices. Fig. 5(a) shows the extracted STT, VCMA, and Joule heating effects as a function of RA for 80 nm devices. The STT effective field decreases rapidly as RA increases, from approximately 10 mT in low-RA devices to below 2 mT in high-RA devices. This behavior can be attributed to the decreasing current density in high-RA devices, which directly weakens the STT effect. Similarly, the Joule heating effective field exhibits a sharper decrease, indicating that it is strongly suppressed in high-RA devices due to lower power dissipation. In contrast, the VCMA effective field remains relatively constant at 20 mT across RAs, highlighting its dependence on the applied gate voltage. **Table 1** summarizes the estimated STT current, STT current density, and increased temperature induced by Joule heating under 1 V gate voltage for three representative RA values in CD = 80 nm devices. This data shows that STT current and current density decrease significantly with increasing RA, while the estimated device temperature drops from 58.8 °C to near room temperature, consistent with reduced Joule heating at higher RA. The temperature rise due to Joule heating was estimated based on the correlation between coercivity and ambient temperature obtained from temperature dependence measurements provided in supplementary materials S8. Fig. 5(b) shows the dependence of effective fields on CD ranging from 50–200 nm for devices with RA = 600 Ω·μm². The effective fields exhibit minimal sensitivity to CD, indicating stable performance with scaling. Minor variations likely arise from RA differences during fabrication. Finally, to assess the potential of SAF-HFL for voltage-gated SOT MRAM, we examine SOT-induced magnetization switching of high RA devices where STT and Joule heating are minimized. In this study, we performed SOT switching measurements under an 40mT in-plane field to achieve deterministic switching. Figure 5(c) shows SOT-induced switching loops with 1ns SOT pulse under different gate voltages, averaged from 50 loops. A clear shift of the switching current is observed while maintaining deterministic transitions. Figure 5(d) summarizes the gate-voltage dependence of the critical SOT current density for both P to AP and AP to P switching from 10 devices, evidencing a systematic and reversible modulation of switching current threshold by the gate voltage.

In summary, we systematically investigated the impact of gate voltage on switching field of MTJs with a SAF free layer. Through micromagnetic simulations and experiments, we demonstrated that localized modulation of anisotropy in the M1 layer governs the switching behavior of the SAF system. We also validated that an analytical model of switching fields developed for single free-layer devices can be effectively extended to SAF, allowing to quantify the voltage-dependent contributions from STT, VCMA, and Joule heating. Our results reveal, in high-RA devices, VCMA effect dominated, leading to a linear voltage dependence of the switching field, while low-RA devices exhibit nonlinear switching behavior due to significant contributions from both STT and Joule heating. The VCMA effect exhibits a consistent linear voltage dependence across RA values, while Joule heating becomes prominent in low-RA devices. Effective field contributions were found limited CD dependence, supporting the scaling potential. Finally, we confirmed that the gate voltage enables reversible control of the critical SOT current, highlighting its potential for voltage-gated SOT MRAM.

This work provided a framework for understanding the roles of STT, VCMA, and Joule heating in SAF free layer devices, enabling systematic optimization of performance and scalability of SOT-MRAM.

**Supplementary Material**

See the Supplementary Material for additional figures, tables, simulation details, and supporting analyses.

**Availability of data**

The data that support the findings of this study are available from the corresponding author upon reasonable request.

**Acknowledgement**

This project has received funding from the European Union's Horizon 2020 research and innovation program under the Marie Skłodowska-Curie grant agreement No 955671. This work was supported by imec's Industrial Affiliation Program on MRAM devices. D.G. and B.V. acknowledge FWO-Vlaanderen for Strategic Basic Research PhD Fellowships (No. 1SHEV24N and No. 1S72225N, respectively). This work has been enabled in part by the NanoIC pilot line. The acquisition and operation are jointly funded by the Chips Joint Undertaking, through the European Union's Digital Europe (101183266) and Horizon Europe programs (101183277), as well as by the participating states Belgium (Flanders), France, Germany, Finland, Ireland and Romania. For more information, visit nanoic-project.eu.